# Mixed Finite Element Analysis of Flexoelectric Response: Exploring Unit Cell Stacking and Strain Gradient Modulation

Arash Kazemi [a], Kshiteej J Deshmukh [b], Susan Trolier-McKinstry [c], Shad Roundy [a]

[a] Department of Mechanical Engineering, University of Utah, Salt Lake City, UT, USA

[b] Department of Mechanical and Aerospace Engineering, University of Houston, Houston, TX 77025

[c] Department of Materials Science and Engineering and Materials Research Institute, The Pennsylvania State University, University Park, PA, USA

* Corresponding author at: Department of Mechanical Engineering, College of Engineering, University of Utah, 1495 E 100 S, Salt Lake City, UT

E-mail address: arash.kazemi@utah.edu

**Abstract:**

Flexoelectricity, a coupling between strain gradients and electric polarization, has attracted significant interest due to its critical role in enhanced effects at small scales and its applicability across a diverse range of materials. Modeling flexoelectricity is challenging, especially in 3D, due to the higher-order equations involved, which require continuity conditions that limit the use of standard finite element methods. This study employs a modified mixed finite element formulation, specifically designed to account for spurious oscillations that can hinder convergence, particularly in large-scale problems solved with iterative solvers. A numerical analysis has been conducted to evaluate the effectiveness of stacking individual flexoelectric unit cells to achieve an enhanced overall flexoelectric response. The study also seeks to determine the feasibility of locally modifying the strain gradient to induce localized enhancements in the flexoelectric effect.

## 1. Introduction

In noncentrosymmetric dielectric crystals, mechanical strain and electric fields are intrinsically coupled, leading to a phenomenon known as piezoelectricity. When a uniform strain is applied to such crystals, it causes a relative displacement between the centers of oppositely charged ions. This displacement generates a net electric dipole moment within the material, linking mechanical strain to electrical polarization linearly. The piezoelectric effect does not occur in dielectric materials with inversion-center symmetry, so piezoelectricity is limited to crystals without this symmetry [1]. However, applying nonuniform strain, or strain gradients, can break the local inversion symmetry within a dielectric. This leads to the flexoelectric effect, where a strain gradient induces polarization in the material.

Although the flexoelectric effect was initially perceived as weak in most solid materials and, consequently, received limited attention at the time of its discovery [2–4], its universality across a wide range of materials and its enhanced significance at small scales—highly advantageous for micro- and nanotechnology—sustained researchers' interest. This interest grew significantly after Cross [5] reported that materials with high dielectric constants could have large flexoelectric coefficients.

The earliest theoretical investigations into flexoelectricity were carried out by Mashkevich and Tolpygo [3,6]. Subsequently, Kogan developed the first phenomenological theory of flexoelectricity and estimated the lower bounds of the flexoelectric coefficients to be on the order of $10^{-9} C/m$ [4]. Following the studies from Tagantsev [7] and Mindlin [8], Sahin and Dost [9] developed a comprehensive and unified continuum framework that encompasses strain gradient elasticity, both direct and converse flexoelectric couplings, as well as the polarization inertia effect. The theoretical formulation of flexoelectricity has been extended through variational models based on free energy minimization, free enthalpy formulations [10–12], and the inclusion of additional physical couplings [13–15]. Furthermore, surface effects have been explored to assess their contributions to the overall flexoelectric response [16].

Progress in the theoretical framework of flexoelectricity has underscored the necessity for experimental verification of newly established theoretical findings. In this context, Ma and Cross [17–19] conducted a systematic investigation of flexoelectricity using three distinct experimental techniques: dynamic cantilever bending, quasi-static four-point bending, and pyramid compression. The first two methods were employed to measure the transverse flexoelectric coefficient, while the pyramid compression technique was utilized to determine the longitudinal flexoelectric coefficient. Their studies revealed a dependency of the flexoelectricity coefficient on the relative permittivity and they reported larger magnitudes ($\approx 10^{-6} C/m$) for ferroelectric perovskites like lead magnesium niobate (PMN), lead zirconate titanate (PZT), and barium strontium titanate (BST).

Classical continuum theory lacks the ability to capture the size effects that are crucial when considering micro- and nanoscale systems, which are particularly relevant in the practical analysis of flexoelectricity. In this regard, size-dependent theories have been proposed using strain-gradient theory [20,21], coupled-stress theory [22,23].

The continuum theory of flexoelectricity is complex for several reasons, including the presence of higher-order partial differential equations, nonlocal effects, and electromechanical couplings within the internal energy. Moreover, the use of various definitions for deformation metrics, along with the requirement for more detailed governing equations and boundary conditions, adds to the difficulty. Due to these complexities, obtaining a closed-form solution for flexoelectricity is generally unfeasible, except in a few simple mechanical models. Consequently, significant efforts in recent years have focused on developing cost-effective and accurate numerical models for simulating flexoelectricity. The fourth-order partial differential equations that emerge from the continuum theory of flexoelectricity require the solution field to exhibit $C^1$-continuity, which limits the applicability of classical finite element methods and presents additional challenges, particularly for three-dimensional applications. Several numerical methods, such as the mixed finite element method [24,25], iso-geometric analysis [26,27], and mesh-free techniques [28], have been utilized to simulate flexoelectricity problems. The mixed finite element method has proven to be the most effective, particularly in handling the imposition of higher-order boundary conditions.

As previously discussed, modeling flexoelectricity is inherently challenging, with the complexity increasing substantially in three-dimensional simulations. Consequently, the number of three-dimensional finite element models capable of accurately simulating flexoelectricity remains limited [28–31]. There remains a need for the development of more accurate models for simulating flexoelectricity. A primary reason for this is the necessity of creating models that can precisely capture the relationship between the measured flexoelectric response and the strain gradient within the structure. This is crucial, as oversimplified analytical models used to estimate the strain gradient distribution are believed to contribute to the overestimation of the flexoelectric constant in experimental results [28,32].

In this study, a custom-developed model based on the mixed finite element formulation is developed to investigate flexoelectricity across various geometrical structures. Specifically, the focus is on examining two primary effects: first, the effectiveness of combining individual flexoelectric unit cells by stacking them both horizontally and vertically to achieve an enhanced overall effect. Second, the feasibility of locally altering the strain gradient, thereby inducing localized enhancement in the flexoelectric response is explored.

For this purpose, in Section 2, the continuum flexoelectric boundary value problem is formulated, accompanied by the necessary Dirichlet and Neumann boundary conditions. In Section 3, a three-dimensional mixed finite element model is presented, including detailed assumptions on discretization and its implementation within FEniCSx [33]. Section 4 offers simulations of a single unit cell (a pyramidal structure) and two-dimensional arrays under varying electrical boundary conditions. Additionally, in this section, the results of introducing voids in the structures and their local effects on strain gradient distributions and the induced electric potential is discussed.

## 2. Constitutive model and boundary-value problem

Various variational principles can be employed to formulate the governing equations and boundary conditions in linear dielectric solids, depending on the selection of state variables used to describe the physical behavior of the system. For the mechanical state variable, the displacement field $u(x)$ is typically selected, while for electrical state variable, common choices include electric polarization $P(x)$, electric displacement $D(x)$, or electric field $E(x)$. The variational form can be derived by minimizing the free energy $\Pi[u(x), P(x)]$, or by optimizing the free enthalpy $\mathcal{H}[u(x), \phi(x)]$ of the system where $\phi(x)$ is the electric potential [32,34–36]. The free enthalpy formulation offers the advantage of automatically satisfying Maxwell–Faraday's law by considering the electric potential, as the electrical state variable [37].

In this section, given the extensive literature on flexoelectric modeling, the derivation of the governing equations used in this work is briefly summarized. Einstein's summation convention is employed, where repeated indices indicate summation over spatial dimensions. Indices following a comma represent spatial derivatives, i.e., $A(x)_{i,j} = \frac{\partial A(x)_i}{\partial x_j}$, where $A(x)$ can be any state variable. For clarity and simplicity, the dependence of state variables on the spatial coordinates $(x)$ is omitted. For a linear flexoelectric solid, the free enthalpy density, incorporating strain gradient effects, can be expressed as follows, where the formulation does not consider the converse flexoelectric effect (i.e., polarization gradient inducing strain) without loss of generality [28].

$$\hbar(\varepsilon_{ij}, E_i, \varepsilon_{jk,l}) = \frac{1}{2} C_{ijkl} \varepsilon_{ij} \varepsilon_{kl} - \frac{1}{2} \kappa_{ij} E_i E_j - \mu_{ijkl} E_i \varepsilon_{jk,l} + \frac{1}{2} h_{ijklmn} \varepsilon_{ij,k} \varepsilon_{lm,n} \qquad (1)$$

where $C_{ijkl}$, $\varepsilon_{ij}$, and $\kappa_{ij}$, are the fourth-order elastic modulus tensor, the second-order strain tensor, and the second-order dielectric tensor, respectively. The electric field $E_i$ is derived from the electric potential, such that $E_i = -\phi_{,i}$. The third term represents the flexoelectric effect, where $\mu_{ijkl}$ is the fourth-order tensor of flexoelectricity. The final term in Eq. (1) accounts for non-local elastic effects or interaction energy of strain-gradients, characterized by the sixth-order strain gradient elasticity tensor $h_{ijklmn}$.

For infinitesimal displacement gradients, the second-order strain tensor can be expressed in terms of the displacement vector components, $u_i$, as

$$\varepsilon_{ij} = \frac{1}{2}(u_{i,j} + u_{j,i}) \tag{2}$$

Considering the enthalpy density in Eq. (1) in the context of dielectrics, the corresponding constitutive relations for the Cauchy stress $\sigma_{ij}$, the higher-order stress $\tau_{ijk}$, and the electric displacement $D_i$, can be expressed as follows [24,38]

$$\sigma_{ij} = \frac{\partial \hbar}{\partial \varepsilon_{ij}} = C_{ijkl}\varepsilon_{kl} = 2\mu\varepsilon_{ij} + \lambda\varepsilon_{kk}\delta_{ij} \tag{3a}$$

$$\tau_{ijk} = \frac{\partial \hbar}{\partial \eta_{ijk}} = h_{ijklmn}\eta_{lmn} - \mu_{lijk}E_l \tag{3b}$$

$$D_i = -\frac{\partial \hbar}{\partial E_i} = \kappa_{ij}E_i + \mu_{ijkl}\eta_{jkl} \tag{3c}$$

where $\mu$ and $\lambda$ are Lamé constants, $\eta_{lmn} = \varepsilon_{mn,l}$ represents the strain gradient, and $\delta_{ij}$ is the Kronecker delta symbol. Note that the higher order stress has contributions from the strain gradient and the electric field. These equations highlight the direct relationships between the mechanical and electrical responses in the material, incorporating both the strain gradient and flexoelectric effects.

The next step involves calculating the total enthalpy, which encompasses both the internal energy and the work performed by or on the system. This calculation is essential for deriving the variational form through the formulation of a corresponding min-max optimization problem. In a physical domain $\Omega$, bounded by the boundary $\partial\Omega$, its decomposition is defined as follows: $\partial\Omega = \partial\Omega_U \cup \partial\Omega_T$, where $\partial\Omega_U$ and $\partial\Omega_T$ correspond to displacement and traction boundaries, respectively. Similarly, the boundary can be expressed as $\partial\Omega = \partial\Omega_D \cup \partial\Omega_R$ for prescribed normal displacement or known higher-order traction boundary condition, $\partial\Omega = \partial\Omega_\phi \cup \partial\Omega_\omega$ for prescribed electric potential or known electric displacement, and $\partial\Omega = \partial\Omega_G \cup \partial\Omega_\Upsilon$ for prescribed electric potential normal derivatives or known higher-order electric displacement boundary condition. These subsets are mutually exclusive, satisfying the conditions $\partial\Omega_U \cap \partial\Omega_T = \partial\Omega_D \cap \partial\Omega_R = \partial\Omega_\phi \cap \partial\Omega_\omega = \partial\Omega_G \cap \partial\Omega_\Upsilon = \emptyset$. The total enthalpy ($\mathcal{H}$) can be written as [34,35]

$$\mathcal{H}(\mathbf{u},\phi) = \int_\Omega \hbar \, d\Omega - \mathcal{W} \tag{4}$$

$\mathcal{W}$ represents the work done by all external forces that for the general case can be written as

$$\mathcal{W} = \int_\Omega f_i \, d\Omega + \int_{\partial\Omega_T} t_i u_i \, dS + \int_{\partial\Omega_R} r_i \partial^n u_i \, dS - \int_{\partial\Omega_\omega} \omega\phi \, dS - \int_{\partial\Omega_\Upsilon} \Upsilon\partial^n\phi \, dS \tag{5}$$

where $\partial^n = \boldsymbol{n}.\boldsymbol{\nabla} = n_i(\frac{\partial}{\partial x_i})$ is the normal derivative, $f_i$ is the body force, and $(t_i, r_i, \omega, \Upsilon)$ are prescribed functions representing traction, double traction, surface charge density, and double charge density respectively [29,30,37]. Using the variational principle, the variation in total enthalpy can be given as,

$$\delta\mathcal{H} = \int_\Omega (-\sigma_{jk,j} + \tau_{ijk,ij} - f_k)\delta u_k \, d\Omega + \int_\Omega (-D_{i,i})\delta\phi \, d\Omega - \int_{\partial\Omega_T}(t_k - (\sigma_{jk} - \tau_{ijk,i})n_j +$$
$$\nabla_j^t(\tau_{ijk}n_i) - \nabla_l^t(n_l)\tau_{ijk}n_i n_j)\delta u_k \, dS - \int_{\partial\Omega_R}(r_k - \tau_{ijk}n_i n_j)\partial^n \delta u_k \, dS + \int_{\partial\Omega_\omega}(\omega +$$
$$D_i n_i)\delta\phi \, dS + \int_{\partial\Omega_\Upsilon}(\Upsilon)\partial^n\delta\phi \, dS \quad (6)$$

where $\nabla_i^t = \frac{\partial}{\partial x_i} - n_i n_j(\frac{\partial}{\partial x_j})$ is the surface gradient. By setting $\delta\mathcal{H} = 0$, the corresponding Euler-Lagrange equations can be derived as follows:

$$\sigma_{jk,j} - \tau_{ijk,ij} + f_k = 0 \quad \text{in } \Omega \quad (7a)$$
$$D_{i,i} = 0 \quad \text{in } \Omega \quad (7b)$$

Accordingly, the associated boundary conditions are defined as follows:

1. Prescribed displacement or known traction:

$$u_k = \tilde{u}_k \text{ on } \partial\Omega_U \quad (8a)$$
$$\tilde{t}_k = (\sigma_{jk} - \tau_{ijk,i})n_i - \nabla_j^t(\tau_{ijk})n_i + \nabla_l^t(n_l)\tau_{ijk}n_i n_j \text{ on } \partial\Omega_T \quad (8b)$$

2. Prescribed normal displacement or known higher-order traction boundary conditions:

$$\partial^n \delta u_k = \tilde{d}_k \text{ on } \partial\Omega_D \quad (9a)$$
$$\tilde{t}_k = \tau_{ijk}n_i n_j \text{ on } \partial\Omega_R \quad (9b)$$

3. Prescribed electric potential or known electric displacement:

$$\phi = \tilde{\phi} \text{ on } \partial\Omega_\phi \quad (10a)$$
$$\tilde{q} = D_i n_i \text{ on } \partial\Omega_\omega \quad (10b)$$

4. Prescribed electric potential normal derivatives or known higher-order electric displacement boundary condition:

$$\partial^n \delta\phi = \tilde{\mathcal{g}} \text{ on } \partial\Omega_G \quad (11a)$$
$$\Upsilon = 0 \text{ on } \partial\Omega_\Upsilon \quad (11b)$$

Note that while the converse flexoelectric effect (polarization gradient giving rise to strain) is not explicitly mentioned, it can be shown using Eq. (4) and integration by parts that the inclusion of both the flexoelectric and converse flexoelectric effects can be accounted for by a term with the same form as the flexoelectric effect [34,36]. Hence, only the flexoelectric term is retained in the above formulation.

## 3. Modified mixed finite elements formulation

The continuum flexoelectricity model developed in the previous section leads to a coupled system of fourth-order partial differential equations as shown in Eq. (7), and the resulting weak form with second order derivatives, which necessitates $C^1$-continuous solution fields. As previously discussed, the $C^1$-continuity requirement, particularly challenging in 3D, limits classical finite element methods, making alternatives like iso-geometric analysis (IGA) and mixed formulations necessary. In mixed formulations mechanical displacement (**u**) and its gradient ($\nabla u$) are treated as independent variables. The mixed variable ($\boldsymbol{\psi}$) is introduced and corresponding constraints that are incorporated into the variational formulation through methods such as using Lagrange multipliers or Nitsche's method [39,40] that can ensure the condition $\nabla u = \boldsymbol{\psi}$. The total enthalpy in Eq. (4) is modified to include the mixed variable and rewritten as,

$$\mathcal{H}^*(\mathbf{u},\phi,\boldsymbol{\alpha},\boldsymbol{\psi}) = \mathcal{H}(\mathbf{u},\phi,\boldsymbol{\psi}) + \int_\Omega \alpha_{jk}(\psi_{jk} - u_{k,j})d\Omega + \int_{\partial\Omega} \zeta_{jk}(\psi_{jk}^t - u_{k,j}^t)dS \qquad (12)$$

here $\psi_{jk}$ represents the displacement gradient, introduced as a new independent variable. The second and third equations impose additional kinematic constraints: $\psi_{jk} = u_{k,j}$ and its tangent component, $\psi_{jk}^t = u_{k,j}^t$ at the bounderies, enforced through the Lagrange multipliers $\alpha_{jk}$, and $\zeta_{jk}$, respectively. By introducing the new variables in $\mathcal{H}^*(\mathbf{u},\phi,\boldsymbol{\alpha},\boldsymbol{\psi})$, the order of the derivatives is reduced, making it possible to utilize $C^0$-continuous elements. It is possible to mathematically demonstrate that the total enthalpy derived in Eq. (12) yields the same boundary value problem as presented in Eqs. (7) – (11). This proof ensures the equivalence of the two formulations. For an in-depth analysis, refer to [24,36].

It is important to recognize that, while Eq. (12) completely enforces all kinematic constraints ($\psi_{jk} = u_{k,j}$ and $\psi_{jk}^t = u_{k,j}^t$), in $C^0$- continuous elements the nodal values ($u_k$) remain continuous across element boundaries, whereas their derivatives ($u_{k,j}$) may exhibit discontinuities. As a result, any constraints that involve these derivatives can only be approximately satisfied within individual elements. At the interfaces between elements, these derivatives may not align perfectly, leading to an approximate enforcement of the constraints. Consequently, it is reasonable to relax the boundary constraint ($\psi_{jk}^t = u_{k,j}^t$) and reformulate Eq. (12) and write the final weak form as

$$\delta \mathcal{H}^*(\mathbf{u}, \phi, \boldsymbol{\alpha}, \boldsymbol{\psi}) \approx \int_\Omega (\sigma_{jk}\delta\varepsilon_{jk} + \tau_{ijk}\delta\psi_{jk,i} - D_i\delta E_i)\, d\Omega + \int_\Omega \delta\alpha_{jk}(\psi_{jk} - u_{k,j})\, d\Omega +$$
$$\int_\Omega \alpha_{jk}(\delta\psi_{jk} - \delta u_{k,j})\, d\Omega - \int_\Omega f_k \delta u_k\, d\Omega - \int_{\partial\Omega_T} t_k \delta u_k\, dS - \int_{\partial\Omega_\omega} \omega\delta\phi\, dS \qquad (13)$$

The above widely used weak form equation can cause spurious oscillations that often result in poor numerical convergence, especially in large cases using iterative solvers. A detailed discussion follows in the next section.

### 3-1. Modified formulation for faster convergence

The formulation introduced in the previous section is of significant interest and has been widely applied in various studies. However, it is susceptible to spurious oscillations, which can lead to poor numerical convergence. This issue has been investigated in elasticity models [41,42], where it has been shown that enforcing constraints via a single Lagrange multiplier per element leads to an averaged enforcement over each element. Consequently, the derivative of the displacement field—computed through direct differentiation and the gradient field—only matches on average and may exhibit local oscillations. As a result, the independently interpolated variable $\psi_{ij}$ can significantly deviate from the displacement field derivative $u_{i,j}$ at specific integration points.

In elasticity and plasticity, penalty terms have been introduced to enforce the kinematic constraint [41,43]. A similar issue has been observed in flexoelectricity problems, particularly when large computational meshes are used and iterative solvers are employed. To mitigate this, a penalty-based strategy has been implemented to improve numerical stability and convergence in flexoelectric simulations. By introducing a penalty factor, denoted as $\wp$, the weak form derived in Eq. (13) was modified in the following manner:

$$\delta \mathcal{T}(\mathbf{u}, \phi, \boldsymbol{\alpha}, \boldsymbol{\psi}) = \delta \mathcal{H}^*(\mathbf{u}, \phi, \boldsymbol{\alpha}, \boldsymbol{\psi}) + \int_\Omega \wp(\psi_{jk} - u_{k,j})(\delta\psi_{jk} - \delta u_{k,j})\, d\Omega \qquad (14)$$

The weak formulation presented above, which incorporates a Lagrange multiplier to enforce the constraint $\nabla \boldsymbol{u} = \boldsymbol{\psi}$, along with a penalty term to enhance the convergence rate, will be discretized in the following section to derive the corresponding finite element equations.

### 4. Discretization

The independent variables in Eq. (14) were discretized over a finite element, denoted as $\Omega^e$. Within each element, the field variables ($\boldsymbol{u}^e, \boldsymbol{\psi}^e, \boldsymbol{\phi}^e, \boldsymbol{\alpha}^e$) are represented by their nodal values ($\boldsymbol{u}^n, \boldsymbol{\psi}^n, \boldsymbol{\phi}^n, \boldsymbol{\alpha}^n$), and interpolation is carried out using $C^0$ shape functions to approximate the variation of these values within the element. The interpolating shape functions used for displacement, displacement gradient, electric potential, and Lagrange multipliers are denoted by $\mathcal{N}_u(x)$, $\mathcal{N}_\psi(x)$, $\mathcal{N}_\phi(x)$, and $\mathcal{N}_\alpha(x)$, respectively, and the field variables are related to the nodal values as shown in Eq. (15). In this formulation, the nodal degrees of freedom are characterized by the following components: three displacement components $u^n$, one electric potential component $\phi^n$, nine components of the displacement-gradient $\psi^n$, and nine components of the Lagrange multipliers $\alpha^n$. Consequently, this leads to a total of 22 unknowns or degrees of freedom per node.

$$u^e(x) = \begin{bmatrix} u_1 \\ u_2 \\ u_3 \end{bmatrix} = \mathcal{N}_u(x)u^n, \psi^e(x) = \begin{bmatrix} \psi_{11} \\ \psi_{12} \\ \psi_{13} \\ \psi_{21} \\ \psi_{22} \\ \psi_{23} \\ \psi_{31} \\ \psi_{32} \\ \psi_{33} \end{bmatrix} = \mathcal{N}_\psi(x)\psi^n, \phi^e(x) = \mathcal{N}_\phi(x)\phi^n, \alpha^e(x) = \begin{bmatrix} \alpha_{11} \\ \alpha_{12} \\ \alpha_{13} \\ \alpha_{21} \\ \alpha_{22} \\ \alpha_{23} \\ \alpha_{31} \\ \alpha_{32} \\ \alpha_{33} \end{bmatrix} = \mathcal{N}_\alpha(x)\alpha^n \qquad (15)$$

To achieve a complete discretization of Eq. (14), it is necessary to compute four additional components: strain ($\varepsilon_{ij}$), displacement gradient ($u_{j,i}$), electric field ($E_i$), and strain-gradient ($\eta_{ijk}$). These quantities are obtained by applying appropriate derivatives and combinations of the shape functions defined in Eq. (15). The calculations follow the standard finite element procedure, ensuring accuracy. Detailed steps for these computations are provided in Appendix B. The resulting expressions for these components in terms of the nodal values of the field variables are given as follows:

$$\varepsilon^e = \mathcal{B}_\varepsilon u^n, \nabla u^e = \mathcal{B}_{\nabla u} u^n, E^e = \mathcal{B}_E \phi^n, \eta^e = \mathcal{B}_\eta \psi^n \qquad (16)$$

With all variables in Eq. (14) either interpolated or calculated using the appropriate shape functions, Eq. (14) can be reformulated into its discrete form (Appendix B), resulting in a system of four corresponding equations for each finite element

$$\begin{bmatrix} \mathbf{K}_{uu} & 0 & \mathbf{K}_{u\psi} & \mathbf{K}_{u\alpha} \\ 0 & \mathbf{K}_{\phi\phi} & \mathbf{K}_{\phi\psi} & 0 \\ \mathbf{K}_{\psi u} & \mathbf{K}_{\psi\phi} & \mathbf{K}_{\psi\psi} & \mathbf{K}_{\psi\alpha} \\ \mathbf{K}_{\alpha u} & 0 & \mathbf{K}_{\alpha\psi} & 0 \end{bmatrix}^e \begin{bmatrix} u \\ \phi \\ \psi \\ \alpha \end{bmatrix}^e = \begin{bmatrix} \mathbf{F}_u \\ \mathbf{F}_\phi \\ 0 \\ 0 \end{bmatrix}^e \qquad (17)$$

in these equations:

$$\begin{aligned} \mathbf{K}_{uu} &= \int_\Omega \mathcal{B}_\varepsilon^T \mathbb{C} \mathcal{B}_\varepsilon d\Omega + p \int_\Omega \mathcal{B}_{\nabla u}^T I \mathcal{B}_{\nabla u} d\Omega \\ \mathbf{K}_{u\psi} &= -p \int_\Omega \mathcal{B}_{\nabla u}^T I \mathcal{N}_\psi d\Omega \end{aligned} \qquad (18)$$

$$\mathbf{K}_{u\alpha} = \int_\Omega \mathcal{B}_{\nabla u}^T \mathbb{I} \mathcal{N}_\alpha d\Omega$$

$$\mathbf{K}_{\phi\phi} = \int_\Omega \mathcal{B}_E^T \mathbb{D} \mathcal{B}_E d\Omega$$

$$\mathbf{K}_{\phi\psi} = \int_\Omega \mathcal{B}_E^T \mathbb{D} \mathcal{B}_\eta d\Omega$$

$$\mathbf{K}_{\psi\psi} = \int_\Omega \mathcal{B}_\eta^T \mathbb{H} \mathcal{B}_\eta d\Omega + p \int_\Omega \mathcal{N}_\psi^T \mathbb{I} \mathcal{N}_\psi d\Omega$$

$$\mathbf{K}_{\psi\alpha} = \int_\Omega \mathcal{N}_\psi^T \mathbb{I} \mathcal{N}_\alpha d\Omega$$

and $\mathbf{K}_{\psi\phi} = \mathbf{K}_{\phi\psi}^T$, $\mathbf{K}_{\alpha u} = \mathbf{K}_{u\alpha}^T$, $\mathbf{K}_{\alpha\psi} = \mathbf{K}_{\psi\alpha}^T$, $\mathbf{K}_{\psi u} = \mathbf{K}_{u\psi}^T$. The right-hand side terms are defined as:

$$\mathbf{F}_u = \int_\Omega \mathcal{N}_u^T \mathcal{F} d\Omega + \int_{\partial\Omega_T} \mathcal{N}_u^T t ds, \quad \mathbf{F}_\phi = \int_{\partial\Omega_\omega} \mathcal{N}_\phi^T \tilde{q} ds \qquad (19)$$

where $\mathcal{F}$, $t$ and $\tilde{q}$ are the body force, the traction vector and the electric displacement on the boundaries. Eq. (17) outlines the elemental formulation of the problem, and after assembling all the elements into a global system we get a linear system of the form $\mathbb{KX} = \mathbb{F}$, where $\mathbb{X}$ represents a finite element space comprising four subspaces: $u$, $\phi$, $\psi$, and $\alpha$. When using multiple finite element spaces to approximate different variables, it is essential to select these spaces carefully to satisfy the stability conditions required for a valid and robust formulation. The proposed method employs four finite element subspaces, and Remark 1 outlines the key criteria for selecting appropriate function spaces in the context of a flexoelectricity problem.

*Remark 1:*

In the context of finite element analysis, the weak forms of the governing equations are derived by multiplying each equation by a corresponding test function, followed by integration over the domain (with second-order terms integrated by parts). This process is then completed by summing the resulting equations. In the Finite Element Method (FEM), the weak form of a differential equation can be directly derived from its variational form (Eq. (14)). The weak formulation defines the appropriate function spaces for both the solution (unknown variables) and the test functions. These function spaces are generally Sobolev spaces. In the discrete setting, the unknowns are expressed through nodal values within each space, and interpolation between these nodal values is achieved using a set of linearly independent functionals.

Previous investigations in the context of strain-gradient elasticity have explored the essential properties that an interpolation scheme must satisfy to ensure convergent results [44]. The essential properties can be summarized as follows

$$sN_{GP} \geq n_\psi \tag{20}$$

where $N_{GP}$ refers to the number of Gauss points used in numerical integration, and $s$ denotes the number of components of the displacement gradient, which is 9 in three-dimensional space. The sufficient conditions for convergence are

$$n_u > n_\psi + sN_{GP} \text{ if } n_\psi \leq sN_{GP} < n_u \tag{21}$$

The variables $n_u$ and $n_\psi$ represent the number of degrees of freedom corresponding to the displacement and displacement gradient, respectively. Based on the criteria in Eqs. (20) and (21), the three displacement components are interpolated with quadratic Lagrangian shape functions, while the electric potential and the nine components of the displacement gradient are interpolated with linear Lagrangian shape functions. Both sets of functions belong to the H¹ Sobolev space. Discontinuous Lagrangian shape functions from the $L^2$ space, specifically piecewise constant elements, are used for the Lagrange multipliers.

## 5. Numerical Simulation

The FEniCS library (an open-source computational platform) is utilized to facilitate the resolution of partial differential equations through the finite element method. It employs Unified Form Language (UFL) for defining weak forms and discretization. For the numerical simulations, the Generalized Minimal Residual (GMRES) method was utilized as the Krylov subspace iterative solver, paired with an Algebraic Multigrid (AMG) preconditioner. This combination within FEniCS proved efficient for managing the large, sparse linear systems resultant from the discretization of the underlying models.

### 5-1. Convergence and validation

To validate the accuracy and convergence of the developed mixed finite element, this section focuses on the flexoelectric response of a flexoelectric spherical shell under compression on its inner or outer surface. Due to the problem's symmetry, as detailed in [29], an analytical solution can be derived. This analytical solution serves as a benchmark against which the proposed mixed finite element formulation can be rigorously compared.

For an isotropic material, Eq. (7) can be simplified by recognizing that the potential $\phi = \phi(r)$ and the radial displacement $u = u(r)$ depend exclusively on the radial coordinate, resulting in the following one-dimensional form [29]:

$$\nabla^2 \phi - \frac{f}{\kappa} \nabla^2 \left( \frac{du}{dr} + \frac{2u}{r} \right) = 0 \tag{22a}$$

$$(1 - \Lambda \nabla^2 - \frac{2\Lambda}{r^2})(\nabla^2 u - \frac{2u}{r^2}) = 0 \qquad (22b)$$

where $\Lambda = \ell^2 + \frac{f^2}{(\lambda+\mu)\kappa}$ and $f = \mu_{11} + 2(\mu_{12} + \mu_{44})$ and $\mu_{11}, \mu_{12}, \mu_{44}$ are the flexoelectric constants, $\kappa$ is the dielectric permittivity and $\ell$ is the material length parameter. Eq. (22) has an analytical solution when the appropriate boundary conditions are applied: a known pressure on the inner surface and zero pressure on the outer surface (see Fig. 1(b)). The resulting variation in electric potential, derived from this analytical solution, is compared with numerical simulations incorporating the material properties outlined in Table 1. This comparison, presented along the central axis of the sphere ($x = y = z$), is illustrated in Fig. 1(a), while the corresponding numerical contour plot is shown in Fig. 1(d). It can be observed that both results agree closely. Next, in order to analyze the convergence of the model, the error was calculated as $e = \frac{\|U_{simulation} - U_{analytical}\|}{\|U_{analytical}\|} \times 100$, where the analytical solution can be obtained solving Eq. (22). The numerical solution was computed using 13727, 37172, 52382, and 55230 elements, yielding total degrees of freedom of 0.30 million, 0.82 million, 1.15 million, and 1.22 million, respectively. The errors associated with each case are presented in Fig. 1(c), where a clear trend of decreasing error with increasing element count is observed.

**Table 1. Material properties** [29]

| Young's modulus | Material length parameter | Flexoelectric constants | Dielectric constant |
|---|---|---|---|
| $E = 139 \, GPa$ | $\ell = 5 \mu m$ | $\mu_{11} = \mu_{12} = 1 \, \mu C/m$<br>$\mu_{44} = 0$ | $\kappa = 1 \, nC/Vm$ |

## 6. Numerical results

Bending structures give rise to strain gradients, which, in turn, may result in electrical polarization. But to achieve very high gradients, sub-micron scale structures are necessary [45]. In case of beam structures, the flexoelectric effect would be significant only in beams with very low thicknesses [11], which might be difficult to achieve in practice with adequate fracture toughness. An alternative means of achieving a heightened strain gradient with larger structures involves compressing truncated pyramids [5].

### 6.1. Single pyramid

Pyramidal structures, when subjected to compression, possess a strain gradient throughout their thickness. This phenomenon could lead to flexoelectric charge separation. In this section, the proposed model is used to investigate flexoelectricity in a pyramidal structure. The flexoelectric properties of pyramidal structures have been examined in previous research through two-dimensional simulations [24,32], with relatively few investigations employing three-dimensional simulations [28,46]. Here, a three-dimensional simulation study is conducted to investigate the performance of a pyramidal structure as a flexoelectric unit cell, which can then be used as input for models of arrays of pyramidal structures.

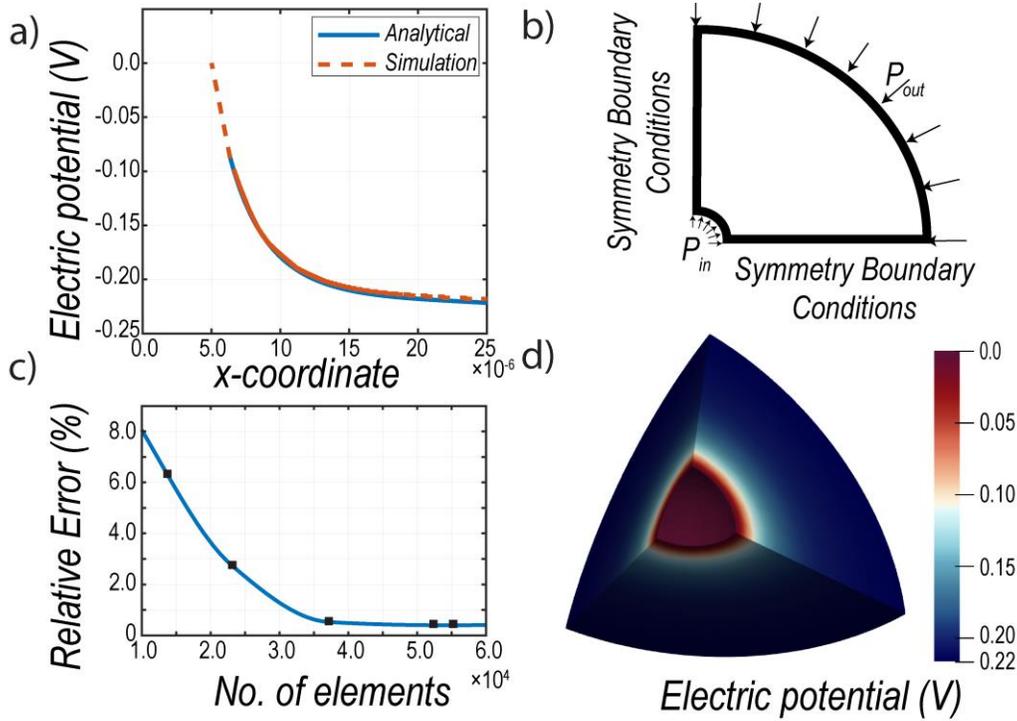

Figure 1: (a) Comparison of the through-thickness electrical potential ($\phi$) from the numerical simulation with the analytical solution. (b) Spherical shell under compression applied to the inner and outer surface. (c) Convergence analysis for different numbers of elements. (d) Contour plot of the electrical potential distribution over the spherical shell.

Fig. 2 shows the 3D simulation results of a truncated pyramid with a height of $h = 0.76mm$ and square top and bottom faces, with lateral dimensions of are $d_1 = 1.13mm$ and $d_2 = 1.72mm$. The material properties utilized in the simulation are provided in Table 2. In simulation, two mechanical boundary conditions were tested: (1) a sliding condition, where a total force of $F = 200N$ is applied uniformly to both the top and bottom faces, and (2) a fixed condition, where the same force is applied to the top face while the bottom is fixed. For the fixed condition, a rectangular base $2.92 \times 2.92 \times 0.05\ mm$ was attached to the bottom face. For the remainder of the paper, the material properties outlined in Table 2 are applied unless stated otherwise. Additionally, the penalty term is set equal to the Young's modulus of the material. The strain distribution presented in Fig. 2 demonstrates that, under sliding boundary conditions, the pyramidal structure experiences greater deformation, leading to increased strain and subsequently higher induced voltage. This observation is consistent with prior research [28], which suggests that flexoelectric effects are more prominent in pyramidal structures subjected to sliding boundary conditions. Additionally, both experimental and numerical studies support the conclusion that pyramidal geometries show strong flexoelectric responses, mainly due to the increased strain gradient under compression [5,28,46,47]. Therefore, the pyramidal shape emerges as a logical candidate for use as a flexoelectric element in stacked transducer configurations.

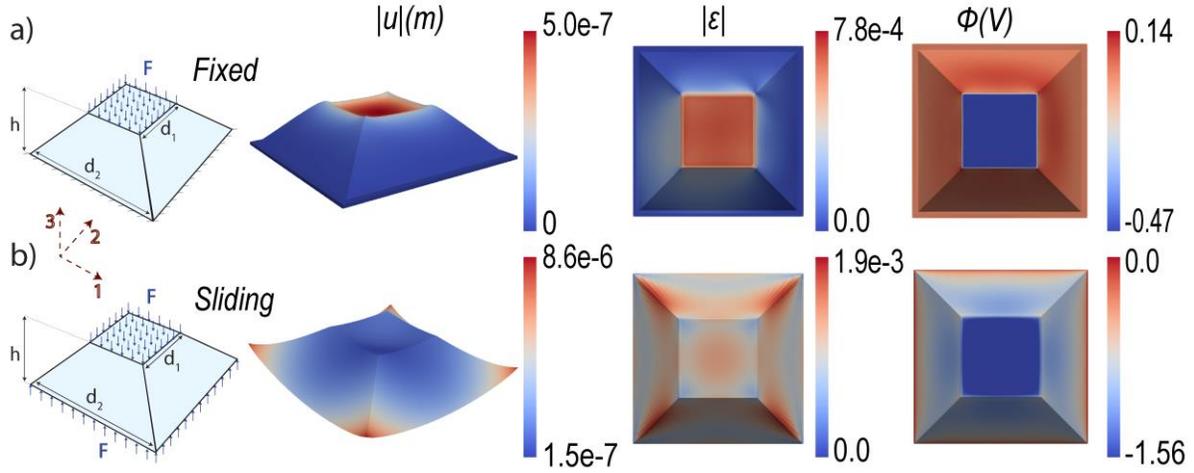

Figure 2: Distribution of the through-thickness mechanical displacement ($|u|$), magnitude of mechanical strain ($|\varepsilon|$), and the electric potential ($\phi$) for a) fixed boundary condition and b) Sliding boundary condition.

In the following section, the performance of a pyramidal structure as a unit cell in a stacked model is examined, in order to explore enhancement of the net flexoelectric response.

Table 2. Material properties (BST)

| Young's modulus | Material length parameter | Flexoelectric constants | Dielectric permittivity |
|---|---|---|---|
| $E = 152\ GPa$ | $\ell = 10\ nm$ | $\mu_{11} = \mu_{12} = 121\ \mu C/m$ <br> $\mu_{44} = 0$ | $\kappa = 146.6\ nC/Vm$ |

## 6.2. 2D arrays: horizontal and vertical

The diagram presented in Fig. 3 illustrates the process of constructing a flexoelectric stack model utilizing a pyramidal structure (Fig. 3(a)) as the fundamental building unit. These units can be assembled in a one-dimensional (1D) array, either horizontally (Fig. 3(b)) or vertically (Fig. 3(c)), and can further be organized into a two-dimensional (2D) array (Fig. 3(d)). To evaluate the performance of the proposed stack model, the flexoelectric response of the one-dimensional (1D) arrays is analyzed in this section. The performance of the 1D horizontal array is identical to that of a single pyramid with its bottom surface fixed and grounded. This was expected, given that neither the electrical nor the mechanical boundary conditions undergo any changes in this configuration. The results obtained for the one-dimensional (1D) vertical array reveal important insights into the model's behavior. Fig. 4 presents the distribution of electric potential across the 1D vertical array, where the bottom surface is maintained at zero potential (grounded) and mechanically fixed, prohibiting any displacement. Additionally, a uniform force of $F = 200N$ is applied to the top surface. Fig. 4(a) demonstrates that the maximum electrical potential is observed at the top surface. Notably, a saddle point is present at the midpoint of the structure where the electric potential changes sign. This is significant because, although the electric potential on the top surface is nearly identical to that of the single-pyramid case with fixed boundary conditions, the configuration has the potential to generate a greater electric potential difference between the top and bottom surfaces compared to the single-pyramid structure. The increased electrical potential difference observed in the upper section of the pyramid can primarily be attributed to the bending occurring in the upper regions of the structure. This phenomenon is clearly illustrated in Fig. 4(b), where the displacement gradients $\psi_{13}$ and $\psi_{23}$ explicitly indicate bending in the upper portion of the structure. It is important to note that, in the case of symmetrical geometry, one would expect the electrical potential on the top surface to be zero. However, this assumption holds only under perfectly symmetric boundary conditions. As previously mentioned, bending occurs in the region near the top surface, whereas the bottom surface remains unaffected by bending due to the fixed boundary condition. Consequently, the mechanical boundary condition is not symmetric. Considering the case of a transducer, placing the two unit cells electrically in series, as shown in the equivalent circuit in Fig. 4(c), would be a poor choice because of the inability to capture the maximum potential difference generated by the structure. Accessing an electrode in between the two pyramids, and electrically placing them in parallel would likely lead to better outcomes as will be shown below. Next, the electrical boundary conditions were modified, and the same structure was simulated with both the bottom and top surfaces grounded. The only mechanical boundary condition applied was a fixed constraint on the bottom surface. Fig. 5(a) illustrates that the maximum electrical potential is nearly the same as in the single pyramid model and is located near the center of the structure. The behavior of the potential is not perfectly symmetric but closely resembles symmetric behavior. This occurs because, although the geometry and electrical boundary conditions are symmetric, the mechanical boundary conditions are not perfectly symmetric and allow greater flexibility near the top surface. It can also be observed that the electric potential profile exhibits an upward shift in the position of the maximum potential in this configuration, with the maximum potential located geometrically above the centerline. This result indicates that a better transducer configuration is to electrically connect the top and bottom surfaces and then measure (or drive) the potential between these two surfaces and the surface between the two pyramids. This configuration effectively wires the two pyramids in parallel, as indicated in the equivalent circuit in Fig. 5(c).

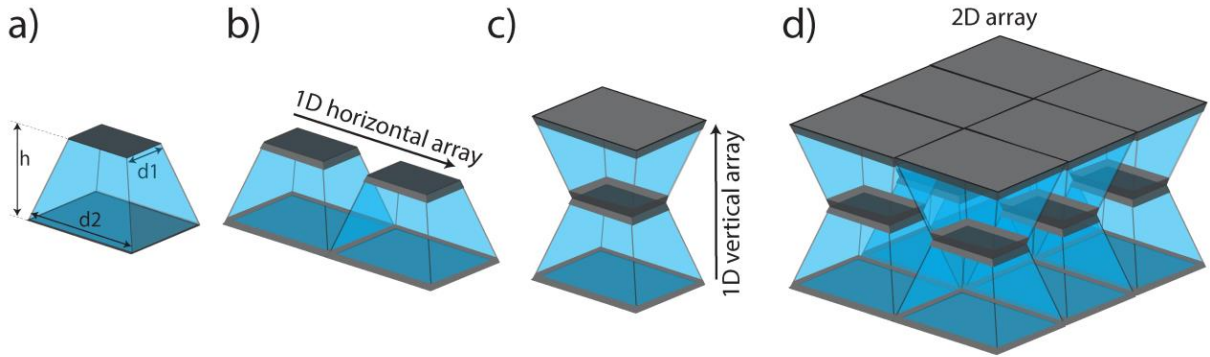

Figure 3: Schematic representation of the construction of a flexoelectric stack model. The basic pyramidal structural unit (a), and the arrangement of these units in a one-dimensional (1D) horizontal (b) and vertical array (c). Formation of a two-dimensional array through combined horizontal and vertical 1D array (d).

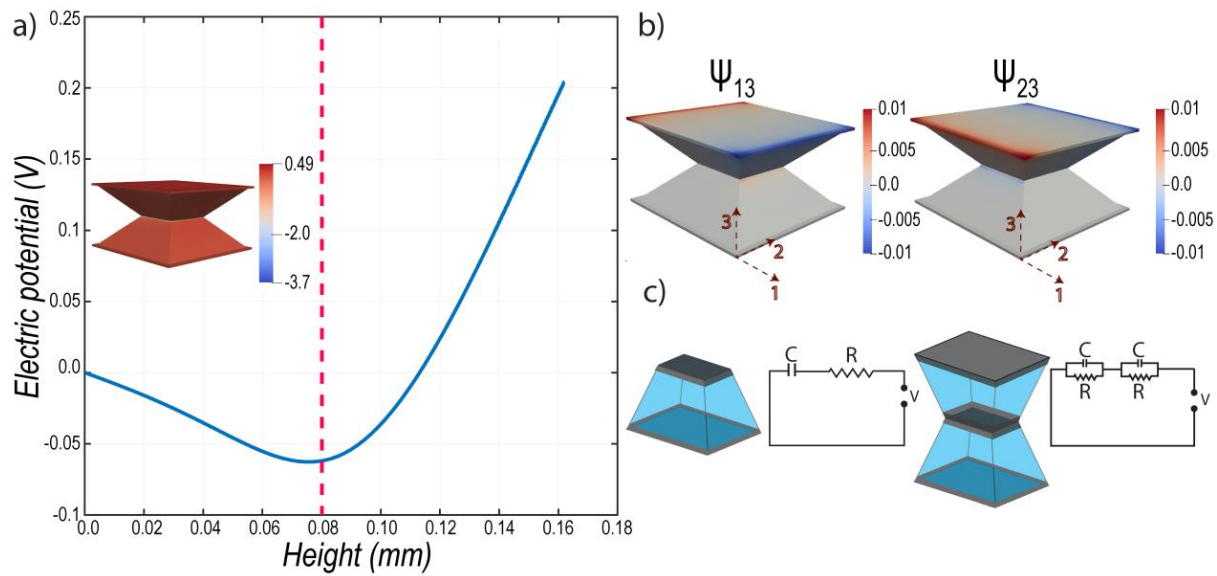

Figure 4: Numerical simulation of a 1D vertical array with the bottom surface grounded (held at zero potential) and mechanically constrained (no displacement). (a) Distribution of electrical potential along the height, (b) displacement gradient illustrating bending in the upper region of the structure, and (c) the equivalent electrical circuit representation of the model.

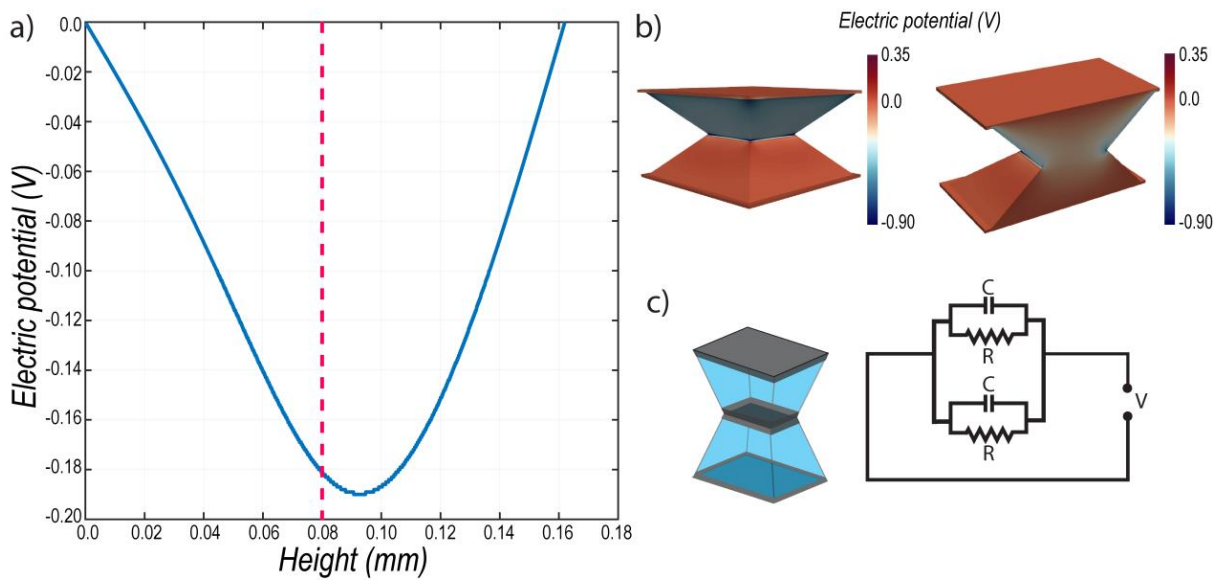

Figure 5: Numerical simulation of a 1D vertical array with the both the bottom and top surfaces grounded (held at zero potential) and the bottom surface mechanically constrained (no displacement). (a) Distribution of electrical potential along the height, (b) contour plots of electric potential, and (c) the equivalent electrical circuit representation of the model.

### 6.3. Flexoelectricity with embedded void

It has been proposed that non-centrosymmetric pores can induce both local and net polarization in non-piezoelectric materials when subjected to uniform stress [48]. In this section, the potential impact of introducing voids on the flexoelectric response of structures is explored. It is reasonable to assume that voids may influence local strain gradients, which could, in turn, affect the induced polarization. To test this assumption, three-dimensional simulations were performed to assess the influence of embedded voids. As an initial case, a pyramidal structure with a cylindrical hole was analyzed. The dimensions of the pyramid are the same as those presented in Fig. 2, with a cylindrical void of $0.1\ mm$ diameter introduced in the $X - Z$ plane at $x = 1mm$ and $z = 0.65\ mm$. The hole extends throughout the entire geometry. Fig. 6(a) presents the electric potential profile plotted along Line A at the top surface of the pyramid. As shown, the presence of the void locally increases the strain gradient within the structure, resulting in a localized enhancement of the induced potential. This indicates that embedded voids exert a localized field effect, with the electric potential diminishing as the distance from the void increases. This local enhancement in electric potential suggests that it may be possible to achieve any desired arbitrary potential pattern.

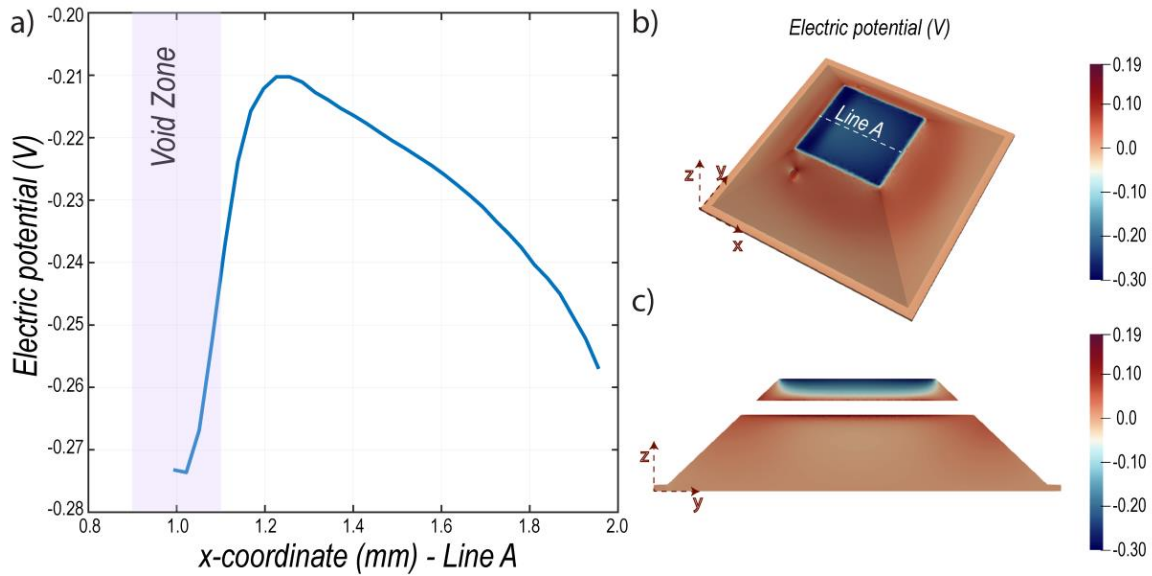

Figure 6: Numerical simulation results for a pyramidal structure containing an embedded void with a fixed bottom surface. (a) Electric potential profile along Line A, located at the top surface of the pyramid. To minimize edge effects, the results are plotted at a distance of $d_1/10$ from the edges, where $d_1 = 1.13 mm$ represents the lateral dimension of the square top face. (b) Contour plots of the electric potential across the structure. (c) Contour plots of the electric potential on a plane passing through the center of the void.

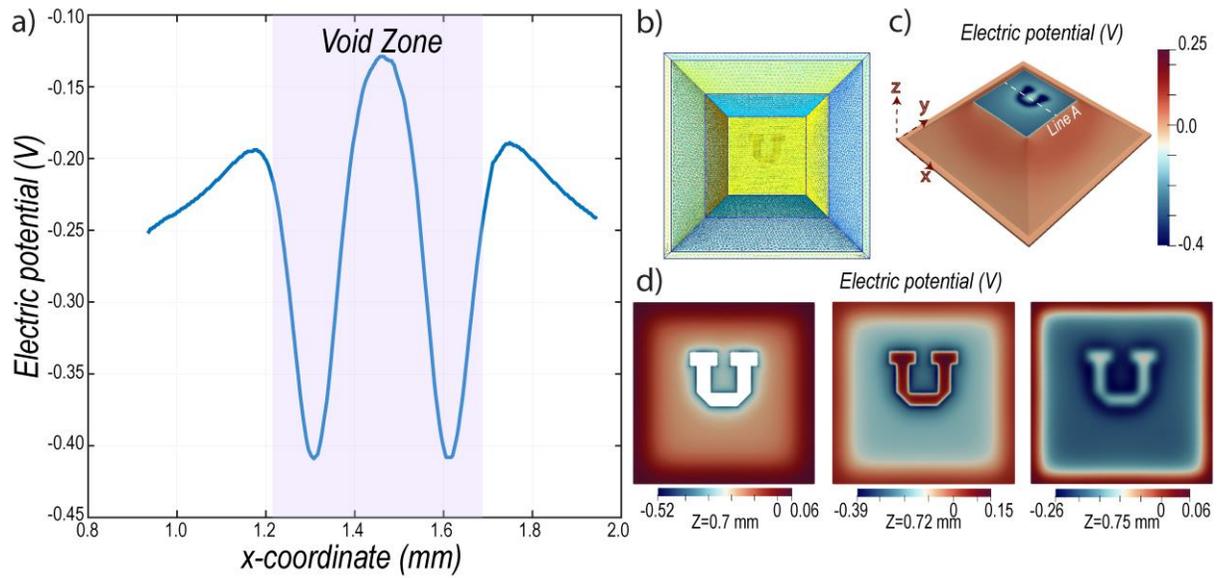

Figure 7: Numerical simulation results for a pyramidal structure with a specifically designed embedded void and a fixed bottom surface. (a) Electric potential profile along Line A, located at the top surface of the pyramid and at $y = 1.6mm$. To minimize edge effects, the results are plotted at a distance of $d_1/30$ from the edges, where $d_1 = 1.13mm$ represents the lateral dimension of the square top face. (b) Partitioned geometry used to generate a denser mesh around the void. (c) Contour plots of the electric potential distribution across the structure. (d) Electric potential profiles at various locations around the void: at $z = 0.7\ mm$ (passing through the void), $z = 0.72\ mm$ (just above the void), and $z = 0.75\ mm$.

To explore the possibility of achieving a specific arbitrary potential profile, Fig. 7 presents the results aimed at producing a U-block pattern on the top surface of a pyramid structure. For this purpose, as shown in Fig. 7(b), the geometry was partitioned to generate a denser mesh around the void. The computational domain was discretized using 919,535 tetrahedral elements. Fig. 7(c) presents the contour plots of the electric potential distribution across the structure, revealing a U-shaped region of elevated potential in comparison to the surrounding areas. Fig. 7(d) presents the electric potential profile at three distinct planes: the first at $z = 0.7\ mm$, which passes through the void; the second at $z = 0.72\ mm$, located just above the void; and the third at $z = 0.75\ mm$. The results suggest the presence of a potential difference between the top surface and the surface above the void, which would generate an electric field in that region. In summary, a comparative analysis of the average induced electric potential for pyramidal structures without voids (Fig. 2) and those with embedded voids (Figs. 6 and 7) reveal minimal variation. The computed average electric potentials on the top surface are -0.520 V for the structure without a void, -0.490 V for the structure with a cylindrical void, and -0.480 V for the structure with a U-block pattern. These findings suggest that while the presence of voids results in a slight reduction in average voltage, the overall impact on the average value remains modest.

## 7. Conclusions

In this study, a general 3D computational model to investigate flexoelectricity was developed. The model is based on a Mixed Finite Element Formulation, which has been modified to address spurious oscillations. These oscillations may arise from discrepancies between the independently interpolated displacement gradient field and the derivative of the displacement field at integration points, leading to poor convergence, particularly in the case of large-scale problems solved using iterative solvers.

Initially, the proposed modified formulation was applied to evaluate the performance of a flexoelectric stack model, where a pyramid serves as the fundamental building unit. Various electrical boundary conditions were explored to assess their potential applications as transducers. The findings indicate that placing two unit cells electrically in series is not ideal, as it fails to capture the maximum potential difference generated by the structure. Instead, introducing an electrode between the two pyramids and configuring them electrically in parallel is likely to yield better results.

Secondly, the potential impact of introducing voids on the flexoelectric response of the structures was explored. The results show that the presence of voids locally amplifies the strain gradient within the structure, leading to a localized enhancement of the induced potential. Furthermore, by precisely engineering the geometric properties of the structure, it is possible to manipulate the strain distribution and, consequently, the resulting flexoelectric response. Although local strain gradient modulation only modestly affects the average response in the studied cases, it opens new avenues for generating novel, tailored potential profiles across the structure.

## CRediT authorship contribution statement

**A. Kazemi:** Writing – original draft, Visualization, Validation, Software, Methodology, Investigation, Formal analysis.

**KJ. Deshmukh:** Writing – review & editing, Supervision, Formal analysis.

**S. Trolier-McKinstry:** Review & editing, Supervision

**S. Roundy:** Writing – review & editing, Supervision, Investigation, Project administration, Formal analysis.

## Declaration of competing interest

The authors declare that they have no known competing financial interests or personal relationships that could have appeared to influence the work reported in this paper.

## Data availability

Data will be made available on request.

## Acknowledgments

This material is based upon work supported by the National Science Foundation under Grant Number (NSF Grant Number 2247453).

## APPENDIX A: material characterization

The free enthalpy density, derived under the assumption of linear flexoelectricity in Eq. (1), is defined by four fundamental material tensors: the elasticity tensor ($C$), dielectric tensor ($\kappa$), flexoelectricity tensor ($\mu$), and strain gradient elasticity tensor ($h$). A detailed, component-wise description of these tensors is provided in the subsequent sections.

### Elasticity tensor:

For isotropic materials, the fourth-order elasticity tensor can be expressed in indicial notation using two independent variables: the Lamé parameters, denoted as ($\mu$ and $\lambda$).

$$C_{ijkl} = \lambda \delta_{ij} + 2\mu \delta_{ik} \delta_{jl} \tag{A-1}$$

The Lamé parameters can be derived in terms of the Young's modulus and Poisson's ratio. In an isotropic homogeneous linear material, the elasticity tensor can be written as

$$\mathbb{C} = \begin{bmatrix} \lambda + 2\mu & \lambda & \lambda & 0 & 0 & 0 \\ \lambda & \lambda + 2\mu & \lambda & 0 & 0 & 0 \\ \lambda & \lambda & \lambda + 2\mu & 0 & 0 & 0 \\ 0 & 0 & 0 & 4\mu & 0 & 0 \\ 0 & 0 & 0 & 0 & 4\mu & 0 \\ 0 & 0 & 0 & 0 & 0 & 4\mu \end{bmatrix} \qquad \text{(A-2)}$$

**Dielectric tensor:**

The isotropic dielectric tensor can be expressed as

$$\kappa_{ij} = \kappa \delta_{ij} \qquad \text{(A-3)}$$

where $\kappa$ is the dielectric constant and for isotropic materials

$$\mathbb{D} = \begin{bmatrix} \kappa & 0 & 0 \\ 0 & \kappa & 0 \\ 0 & 0 & \kappa \end{bmatrix} \qquad \text{(A-4)}$$

**Flexoelectricity tensor:**

For materials with cubic symmetry, the fourth-order flexoelectric tensor can be described in terms of three independent variables: the longitudinal ($\mu_{11}$), transversal ($\mu_{12}$), and shear parameters ($\mu_{44}$).

$$\mu_{ijkl} = (\mu_{11} - \mu_{12} - 2\mu_{44})\,\delta_{ijkl} + \mu_{12}\delta_{ij}\delta_{kl} + \mu_{44}(\delta_{ik}\delta_{jl} + \delta_{il}\delta_{jk}) \qquad \text{(A-5)}$$

and $\delta_{ijkl}$ is one for all indices equal and zero otherwise. The cubic flexoelectric tensor can be written in terms of the three independent parameters

$$\mathbb{F} = \begin{bmatrix} \mu_{11} & 0 & 0 & \mu_{12} & 0 & 0 & \mu_{12} & 0 & 0 & 0 & \mu_{44} & 0 & 0 & 0 & \mu_{44} & 0 & 0 & 0 \\ 0 & \mu_{12} & 0 & 0 & \mu_{11} & 0 & 0 & \mu_{12} & 0 & \mu_{44} & 0 & 0 & 0 & 0 & 0 & 0 & \mu_{44} & 0 \\ 0 & 0 & \mu_{12} & 0 & 0 & \mu_{12} & 0 & 0 & \mu_{11} & 0 & 0 & 0 & \mu_{44} & 0 & 0 & \mu_{44} & 0 & 0 \end{bmatrix} \qquad \text{(A-6)}$$

**Strain gradient elasticity tensor:**

By simplifying the general strain gradient elasticity model [20] for the case of isotropic homogeneous linear material, the sixth-order strain gradient elasticity tensor can be expressed in terms of the Lamé parameters ($\lambda$, $\mu$) and a mechanical length scale parameter ($\ell$).

$$h_{ijklmn} = \left(\lambda \delta_{ij}\delta_{lm} + 2\mu \delta_{il}\delta_{jm}\right) \ell^2 \, \delta_{kn} \qquad \text{(A-7)}$$

which can be reformulated as

$$\mathbb{H} = \begin{bmatrix} \ell^2 \mathbb{C} & 0 & 0 \\ 0 & \ell^2 \mathbb{C} & 0 \\ 0 & 0 & \ell^2 \mathbb{C} \end{bmatrix} \tag{A-8}$$

## APPENDIX B: detailed discretization of the finite element model

To derive the stiffness matrix associated with Eq. (14), as outlined in Eq. (15), the independent variables $u$, $\phi$, $\psi$, and $\alpha$ must be interpolated using the corresponding shape functions $\mathcal{N}_u$, $\mathcal{N}_\psi$, $\mathcal{N}_\phi$, $\mathcal{N}_\alpha$, respectively. Nonetheless, Eq. (14) is also dependent on the strain ($\varepsilon_{ij}$), the displacement gradient ($u_{j,i}$), the electric field ($E_i$), and the strain-gradient ($\eta_{ijk}$), all of which require calculation. The strain tensor and the displacement gradient can be directly obtained from the displacement field as follows:

$$\boldsymbol{\varepsilon}^e = \begin{bmatrix} \varepsilon_{11} \\ \varepsilon_{22} \\ \varepsilon_{33} \\ 2\varepsilon_{12} \\ 2\varepsilon_{23} \\ 2\varepsilon_{13} \end{bmatrix} = \begin{bmatrix} \frac{\partial}{\partial x_1} & 0 & 0 \\ 0 & \frac{\partial}{\partial x_2} & 0 \\ 0 & 0 & \frac{\partial}{\partial x_3} \\ \frac{1}{2}\frac{\partial}{\partial x_2} & \frac{1}{2}\frac{\partial}{\partial x_1} & 0 \\ 0 & \frac{1}{2}\frac{\partial}{\partial x_3} & \frac{1}{2}\frac{\partial}{\partial x_2} \\ \frac{1}{2}\frac{\partial}{\partial x_3} & 0 & \frac{1}{2}\frac{\partial}{\partial x_1} \end{bmatrix} \mathcal{N}_u u^n = \mathcal{B}_\varepsilon u^n, \quad \nabla u^e = \begin{bmatrix} \frac{\partial \mathcal{N}_u}{\partial x_1} \\ \frac{\partial \mathcal{N}_u}{\partial x_2} \\ \frac{\partial \mathcal{N}_u}{\partial x_3} \end{bmatrix} = \mathcal{B}_{\nabla u} u^n \tag{B-1}$$

The electric field $E_i$ is derived from the electric potential $\phi$ according to the relation $E_i = -\phi_{,i}$. Therefore, the electric field can be expressed as:

$$\boldsymbol{E}^e = \begin{bmatrix} -\frac{\partial \mathcal{N}_\phi}{\partial x_1} \\ -\frac{\partial \mathcal{N}_\phi}{\partial x_2} \\ -\frac{\partial \mathcal{N}_\phi}{\partial x_3} \end{bmatrix} = \mathcal{B}_E \phi^n \tag{B-2}$$

The final variable that needs to be derived is the strain gradient, which is obtained by taking the first derivative of $\psi$, as follows:

$$\boldsymbol{\eta}^e = \begin{bmatrix} \frac{\partial \mathbb{N}}{\partial x_1} \\ \frac{\partial \mathbb{N}}{\partial x_2} \\ \frac{\partial \mathbb{N}}{\partial x_3} \end{bmatrix} = \mathcal{B}_\eta \psi^n \tag{B-3}$$

where $\mathbb{N}$ is

$$\mathbb{N} = \begin{bmatrix} 1 & 0 & 0 & 0 & 0 & 0 & 0 & 0 & 0 \\ 0 & 0 & 0 & 0 & 1 & 0 & 0 & 0 & 0 \\ 0 & 0 & 0 & 0 & 0 & 0 & 0 & 0 & 1 \\ 0 & \frac{1}{2} & 0 & \frac{1}{2} & 0 & 0 & 0 & 0 & 0 \\ 0 & 0 & 0 & 0 & 0 & \frac{1}{2} & 0 & \frac{1}{2} & 0 \\ 0 & 0 & \frac{1}{2} & 0 & 0 & \frac{1}{2} & 0 & 0 & 0 \end{bmatrix} \mathcal{N}_\psi \qquad \text{(B-4)}$$

By interpolating or computing all variables in Eq. (14) using the appropriate finite element shape functions, Eq. (14) can be transformed into its discrete form as

$$\begin{aligned}
& \delta \boldsymbol{u}^T \int_\Omega \boldsymbol{\mathcal{B}}_\varepsilon^T \mathbb{C} \boldsymbol{\mathcal{B}}_\varepsilon d\Omega \boldsymbol{u} + \delta \boldsymbol{\psi}^T \int_\Omega \boldsymbol{\mathcal{B}}_\eta^T \mathbb{H} \boldsymbol{\mathcal{B}}_\eta d\Omega \boldsymbol{\psi} - \delta \boldsymbol{\phi}^T \int_\Omega \boldsymbol{\mathcal{B}}_E^T \mathbb{F} \boldsymbol{\mathcal{B}}_\eta d\Omega \boldsymbol{\psi} - \\
& \delta \boldsymbol{\psi}^T \int_\Omega \boldsymbol{\mathcal{B}}_\eta^T \mathbb{F}^T \boldsymbol{\mathcal{B}}_E d\Omega \boldsymbol{\phi} + \delta \boldsymbol{\alpha}^T \int_\Omega \boldsymbol{\mathcal{N}}_\alpha^T I \boldsymbol{\mathcal{N}}_\psi d\Omega \boldsymbol{\psi} - \delta \boldsymbol{\alpha}^T \int_\Omega \boldsymbol{\mathcal{N}}_\alpha^T I \boldsymbol{\mathcal{B}}_{\nabla u} d\Omega \boldsymbol{u} + \\
& \delta \boldsymbol{\psi}^T \int_\Omega \boldsymbol{\mathcal{N}}_\psi^T I \boldsymbol{\mathcal{N}}_\alpha d\Omega \boldsymbol{\alpha} - \delta \boldsymbol{u}^T \int_\Omega \boldsymbol{\mathcal{B}}_{\nabla u}^T I \boldsymbol{\mathcal{N}}_\alpha d\Omega \boldsymbol{\alpha} - \delta \boldsymbol{\phi}^T \int_\Omega \boldsymbol{\mathcal{B}}_E^T \mathbb{D} \boldsymbol{\mathcal{B}}_E d\Omega \boldsymbol{\phi} + \\
& \delta \boldsymbol{u}^T \int_\Omega \mathcal{p} \boldsymbol{\mathcal{B}}_{\nabla u}^T I \boldsymbol{\mathcal{B}}_{\nabla u} d\Omega \boldsymbol{u} - \delta \boldsymbol{u}^T \int_\Omega \mathcal{p} \boldsymbol{\mathcal{B}}_{\nabla u}^T I \boldsymbol{\mathcal{N}}_\psi d\Omega \boldsymbol{\psi} - \delta \boldsymbol{\psi}^T \int_\Omega \mathcal{p} \boldsymbol{\mathcal{N}}_\psi^T I \boldsymbol{\mathcal{B}}_{\nabla u} d\Omega \boldsymbol{u} + \\
& \delta \boldsymbol{\psi}^T \int_\Omega \mathcal{p} \boldsymbol{\mathcal{N}}_\psi^T I \boldsymbol{\mathcal{N}}_\psi d\Omega \boldsymbol{\psi} = \delta \boldsymbol{u}^T \int_\Omega \boldsymbol{\mathcal{N}}_u^T \boldsymbol{\mathcal{F}} d\Omega + \delta \boldsymbol{u}^T \int_{\partial\Omega} \boldsymbol{\mathcal{N}}_u^T \boldsymbol{t} ds + \delta \boldsymbol{\phi}^T \int_{\partial\Omega} \boldsymbol{\mathcal{N}}_\phi^T \boldsymbol{q} ds
\end{aligned} \qquad \text{(B-5)}$$

where $\boldsymbol{\mathcal{F}}$, $\boldsymbol{t}$ and $\boldsymbol{q}$, represent the body force, the traction vector, and the electric displacement on the boundaries, respectively. The above equation will be used for the derivation of the stiffness matrix.